\documentclass[reprint,a4paper,10pt,twocolumn,aps,prb,nopacs,superscriptaddress,longbibliography]{revtex4-2}
\usepackage[top=0.8in,bottom=0.8in,left=0.64in,right=0.64in]{geometry}
\usepackage{csquotes}
\usepackage{sidecap}
\usepackage{ragged2e}
\usepackage[utf8]{inputenc}
\usepackage{fancyhdr}
\usepackage{amsmath}
\usepackage{amsfonts}
\usepackage{amssymb}
\usepackage{subcaption}
\usepackage[nottoc]{tocbibind}
\usepackage{float}
 \usepackage[table]{xcolor}
\usepackage{tabularray}
\usepackage[english]{babel}
\usepackage[utf8]{inputenc}
\usepackage{newunicodechar}
\newunicodechar{≤}{\ensuremath{\leq}}
\usepackage{caption}
\usepackage{graphicx}
\usepackage{braket}
\usepackage{tikz}
\usepackage{color,soul}
\usepackage[
    colorlinks=True,
    linkcolor=blue,
    filecolor=magenta,
    urlcolor=blue,
    citecolor=blue,]{hyperref}
\begin{document}
	\title{\Large{\textbf{Evolution of ferrimagnetism against Griffiths singularity in Calcium Ruthenate}}}
	\author{Pooja}
	\email{kpooja@iitk.ac.in}
	\affiliation{Department of Physics, Indian Institute of Technology, Kanpur 208016, India}
	\author{Sachindra Nath Sarangi}
\affiliation{Institute of Physics, Bhubaneswar 751005, India}
   \author{D. Samal}
\affiliation{Institute of Physics, Bhubaneswar 751005, India}
\affiliation{Homi Bhabha National Institute, Anushaktinagar, Mumbai 400094, India}
   \author{Chanchal Sow}
   \email{chancahal@iitk.ac.in}
\affiliation{Department of Physics, Indian Institute of Technology, Kanpur 208016, India}
\email{chanchal@iitk.ac.in}	
  \date{\today}
	\begin{abstract}
The magnetism in the correlated metal CaRuO$_3$ is enigmatic as it is poised near a triple point among the ferromagnetic, antiferromagnetic, and paramagnetic ground states. Here we report a detailed work on structural, spectroscopic, magnetic, and transport properties in CaRu$_{1-x}$Cr$_x$O$_3$. We find that Cr doping reduces the orthorhombicity in CaRuO$_3$. Surprisingly, a tiny (x = 0.01) amount of Cr-doping drives the magnetic ground state from \enquote{paramagnetic-like} to ferrimagnetic. Slightly higher Cr-doping (x = 0.05) results formation of magnetic clusters which gives rise to Griffiths singularity and power law divergence in magnetic susceptibility. The magnetism in CaRu$_{1-x}$Cr$_x$O$_3$ is explained in terms of \enquote{seven atom} ferrimagnetic clusters. Electrical transport shows a gradual evolution of a non-metallic state upon Cr-doping. In particular, for x $\geq$ 0.1, the temperature-dependent resistivity follows Mott-VRH conduction. The XPS study also supports significant role of disorder and electron correlation which effectively reduces the itinerant character of electrons. Finally, a new T-x phase diagram is constructed depicting the evolution of electronic and magnetic state in CaRu$_{1-x}$Cr$_x$O$_3$.
	\end{abstract}
	\maketitle
	\section{Introduction}
  The low-temperature magnetism and electronic transport in strongly correlated systems continue to surprise to date. Ruthenates are members of the 4d transition metal oxide (TMO) family and exhibit a variety of magnetic ground states, starting from itinerant ferromagnetism (FM) in SrRuO$_3$ \cite{longo1968magnetic}, Mott-insulator in Ca$_2$RuO$_4$ \cite{nakatsuji19972}, semi-conducting metal in La$_2$RuO$_5$ \cite{khalifah2002orbital} and antiferromagnetic (AFM) polar metal in Ca$_3$Ru$_2$O$_7$ \cite{naoki2021magnetoentropic}. The magnetic ground state of CaRuO$_3$ has been labeled as paramagnet (PM) \cite{kiyama1998specific}, FM \cite{mukuda1999spin}, and AFM \cite{longo1968magnetic} by various groups. The main reason for this controversy lies in the structural distortion, which originates primarily due to Ru/O vacancy/excess. CaRuO$_3$ exhibits an orthorhombic crystal structure with space group $Pnma$ (62), where a central Ca atom is surrounded by corner-sharing RuO$_6$ octahedra \cite{bensch1990structure}. External perturbation such as vacancy, chemical doping, pressure, and strain induces significant octahedral distortions \cite{he2001disorder,kojitani2007post,mazin1997electronic}. The magnetic ground state of CaRuO$_3$ is prone to these perturbations since it alters the Ru-O-Ru bond angles, and bond lengths which consequently alters the crystal field splitting and exchange interaction \cite{dang2015electronic}. The work by $\emph{Nguyen et al.}$  \cite{PhysRevMaterials.4.034407} reports the evolution of itinerant ferromagnetism with varying Ru-O-Ru bond angle in Ca$_x$Sr$_{1-x}$RuO$_3$. The mismatch in the ionic radius of  Sr$^{2+}$ (0.131 nm) and Ca$^{2+}$ (0.118 nm) causes the change in the octahedral bond angle. In thin films, tensile strain induces ferromagnetism in CaRuO$_3$ \cite{tripathi2014ferromagnetic, chen2010microstructure}. DFT calculation finds that \enquote{orthorhombic CaRuO$_3$} lies on the border of three distinct magnetic ground states namely FM, C-type AFM, and PM while \enquote{cubic CaRuO$_3$} always goes into FM ground states \cite{zhang2022density}. Recent work by $\emph{Shen et al.}$ \cite{shen2021emergent} demonstrated that the protonated CaRuO$_3$ exhibits the non-Fermi liquid (NFL) to Fermi liquid (FL) caused by the modulation of density of state(DOS). These findings suggest that the electronic, magnetic ground state of CaRuO$_3$ is delicate.\par
The study of chemical doping in the correlated system is very important as it acts as a random potential \cite{kim2005metal}. Under such strong random potential, the electron's wave function gets localized, leading to the metal-to-insulator transition (MIT) \cite{anderson1958absence}. In a correlated system, MIT can be explained by the Mott-Hubbard model \cite{vollhardt2020dynamical}, described as-
\begin{equation}
	H = -t\sum_{<i,j>\sigma}c_{i\sigma}^{\dagger}c_{j\sigma} +  U\sum_i n_{i\uparrow}n_{i\downarrow} + \sum_{i\sigma}\epsilon_i n_{i\sigma}
	\label{equ:1}
\end{equation}
where $c_{i\sigma}^{\dagger}$ and $ c_{j\sigma}$ are the creation and annihilation operators for spin-$\sigma$ electrons at $i$ and $j$ site respectively. The first term is kinetic energy where $t$ is the hopping integral for nearest-neighbor ($nn$) electrons. The second term is correlation energy, and $U$ is on-site Coulomb repulsion. The third term is disorder potential energy where $\epsilon_i$ is random site potential. In Eq-\ref{equ:1} the strength of $t$, $U$, and $\epsilon_i$ mutually decide the fate of the transport behavior in correlated systems \cite{inoue1995systematic,lee2002electron}. A more specific model could be using dynamical mean-field theory \cite{georges1996dynamical}, which can describe about the magnetic correlations as well. Chemical doping at the Ru-site has been studied with various 3d transition metal ions (Mn$^{4+}$, Fe$^{3+}$, Co$^{2+}$, Ni$^{2+}$, Cr$^{3+}$, and Cu$^{2+}$)  \cite{he2001effect,taniguchi2009ferromagnetism,maignan2006ferromagnetism,durairaj2006highly,taniguchi2008crystallographic}. 
CaRu$_{1-x}$M$_x$O$_3$ (M = Ti, Fe, Cu) possess slow magnetization dynamics with short-range magnetic clusters  \cite{felner2002magnetic,bradaric2001metal,bradaric2018anomalous,he2001effect}.  While in transport metal to insulator transition is observed due to disorder induced electron localization effect \cite{cao1996itinerant,singh2022quantum}. Mn-doped CaRuO$_3$ exhibits enhanced ferromagnetism with large magnetoresistance \cite{giri2014co,liu2014enhanced}. The doping also introduces disorder and competing interactions which lead to slow magnetization dynamics. In certain cases, due to the phase inhomogeneity and quenched disorder above the Curie temperature Griffiths Phase (GP) appears. Originally GP was discovered in diluted Ising ferromagnets \cite{griffiths1969nonanalytic}. The random distribution of nearest neighbour exchange bonds generates the intermediate phase between the FM and PM. The temperature regime between T$_c$ and T$_G$ (GP temperature) is very interesting where the system neither exhibits pure PM nor long-range FM ordering. This intermediate GP regime is microscopically characterised by cluster induced by disordered state \cite{neto1998non}. Experimentally the GP is characterised by anomalies in the magnetic susceptibility, specific heat or transport properties \cite{salamon2002colossal,neto1998non,de1998evidence,yamamoto2020electronic,mello2020griffiths,andrade2009electronic}. In correlated systems such as Manganites, Cuprates and Ruthenates, the quenched disorder arises owing to the mismatch in ionic radius while doping, mixed valency in the magnetic ions, octahedral tilt and distortions etc. The existence of the cluster distribution of a GP has been claimed to be responsible for large magnetoresistance in LaMnO$_3$ \cite{salamon2002colossal}, and  Mn-doped SrRuO$_3$ \cite{banerjee2001enhanced}.\\
Among these ions Cr-doping in CaRuO$_3$ is the most interesting case as it induces stronger ferromagnetism than others \cite{maignan2006ferromagnetism}. Cr-doping also results squeezing of the lattice with highly anisotropic and complex magnetic behaviour \cite{durairaj2006highly,maignan2006ferromagnetism} , including disorder induced ferromagnetism \cite{he2001disorder}. However, the saturation magnetic moment is five times smaller than the iso-structural SrRuO$_3$. While the transport study finds a gradual evolution of insulating nature with Cr doping in CaRuO$_3$. The origin of enhanced magnetism and the evolution of such insulating behavior need a systematic study. This article investigates structure, magnetism, and transport properties in CaRu$_{1-x}$Cr$_x$O$_3$ (0 $\geq$ x $\geq$ 0.15) and finds interesting evolution of \enquote{ferrimagnetic phase} against \enquote{Griffiths phase}.
\section{Sample preparation and experimental details-}
Cr-doped CaRuO$_3$ (CaRu$_{1-x}$Cr$_x$O$_3$ denoted as CRCO ) samples with x = 0, 0.01, 0.05, 0.10, and 0.15 are synthesized via the solid-state reaction method. High purity CaCO$_3$ ($\geq$ 99.95\%), RuO$_2$ ($\geq$ 99.9\%), and Cr$_2O_3$ ($\geq$ 99.9\%) powders are mixed in a stoichiometric ratio. The mixture is palletized into 15 mm pallets at 50 kPa/$cm^2$ pressure followed by sintering for 48 h at 900 - 1300 $^{\circ}$C with five intermediate grindings. 
   The phase purity of CRCO samples is verified by high-resolution powder X-ray powder diffraction (PAN analytical Empyrean Cu K$\alpha$ radiation). The atomic percentage of Ru and Cr content of various samples (Tabulated in Table-I) is carried out by the electron probe microscopy analyzer (EPMA) coupled with JXA-8230; JEOL. The oxidation states of Cr, Ru, and Ca are examined by X-ray photoelectron spectroscopy (XPS) using PHI Versa Probe-II equipped with a monochromatic Al K$\alpha$ (h$\nu$ = 1486.6 eV). DC magnetization and the magneto-transport study are carried out in the temperature range of 2 - 300 K and magnetic field 0 - 6 T using commercial quantum design MPMS (MPMS-XL) and PPMS  respectively.
 \begin{table}
\begin{center}
\caption{\justifying{\small{Results of EPMA : where x$_{nom}$ and x$_{det}$ are nominal and determined value of Cr concentration respectively. Ru\% and Cr\% are atomic percentage values in CRCO}}}
\begin{tblr}{colspec={|ccccc|},hlines}
Sample & x$_{nom}$&x$_{det}$ &Ru\%&Cr\%   \\ 
CRO &0 & 0  & 99.5&0   \\
CRCO5 &0.05  & 0.056 (1) &97.5& 5.6  \\ 
CRCO10&0.10 &0.132 (2) &95.6 &13.2 \\ 
CRCO15&0.15 &0.154 (1) &85.5 & 15.4 \\ 
\end{tblr}
\end{center}
\end{table}
\section{Experimental RESULTS-}
\subsection{Structural analysis}
\begin{figure}
\begin{flushright}
\includegraphics[width=\linewidth]{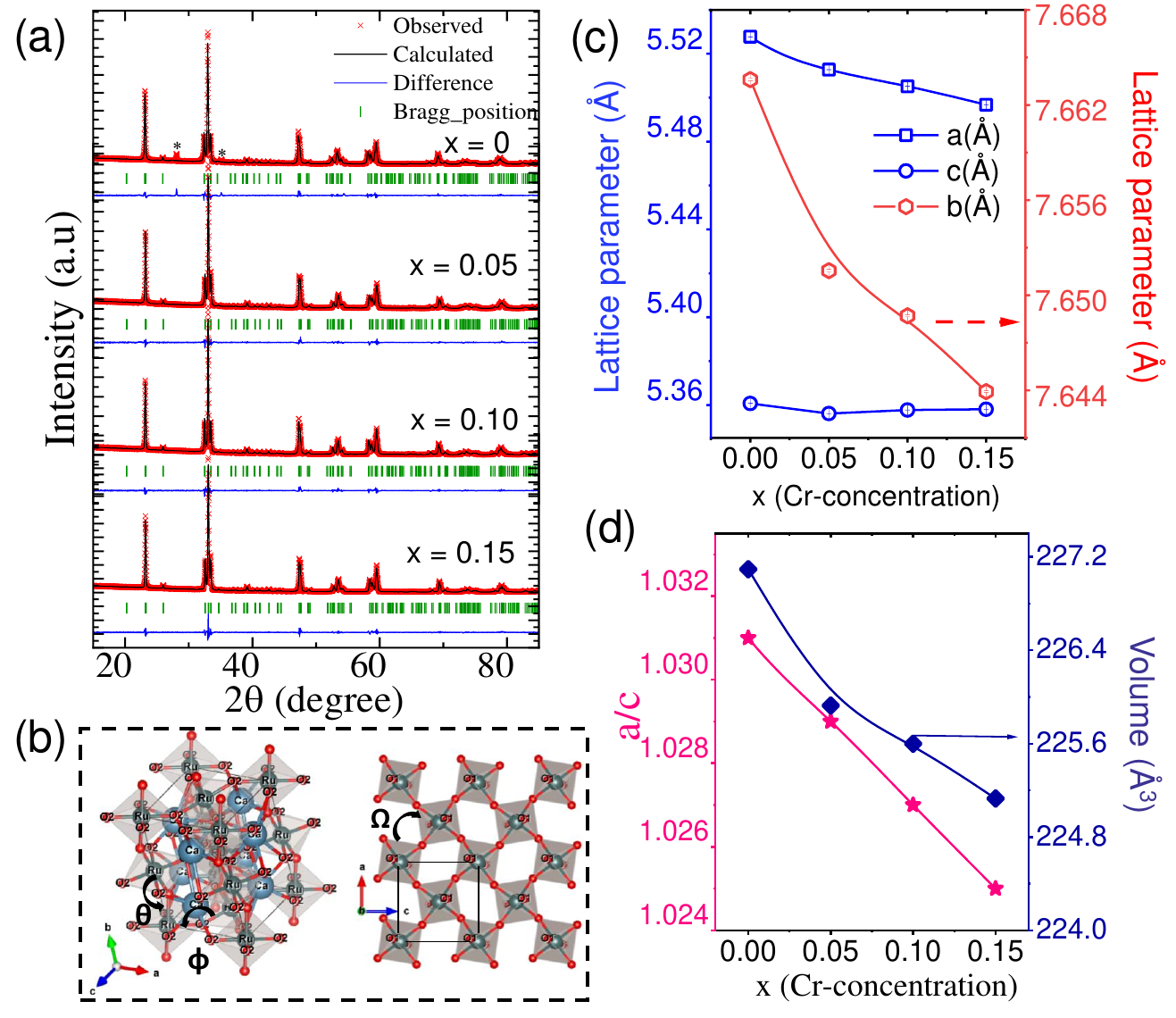} 
 \caption{\justifying{\small{(a) XRD pattern of CRCO samples at x = 0, 0.05, 0.10, and 0.15, along with the Rietveld refinement. (b) 3D and 2D views of the orthorhombic crystal structure of CRO (using \href{https://jp-minerals.org/vesta/en/}{VESTA} visualizing program \cite{momma2008vesta}) with corner-sharing RuO$_6$ octahedra, O1 (Wyckoff - 4c) and O2 (Wyckoff - 8d) are two different site oxygen, $\theta$ and $\phi$ are the Ru-O1-Ru (along b-axis) and Ru-O2-Ru (along a, and c axes) angles respectively. The in-plane rotation angle among three O2 ions is shown by $\Omega$. (c) Variation of lattice constants a, b, and c (d) Orthorhombicity factor (a/c) and unit cell volume, is plotted against Cr-concentrations.}}}
\label{fig:C0}
\end{flushright}
\end{figure}

The crystal structure of CaRuO$_3$ is very sensitive as a small amount of Cr-doping at the Ru-site (Wyckoff - 4a) changes the crystal structure considerably. Fig-\ref{fig:C0}(a) shows the Rietveld-refined (using \href{http://www.ill.eu/sites/fullprof/}{FULLPROF} \cite{ritter2011neutrons}) X-ray diffraction patterns of the CRCO samples without any impurity phases. The refined crystal structure is found to be orthorhombic (a $\neq$ b $\neq$ c, $\alpha$ = $\beta$ = $\gamma$ = 90) with space group $Pnma$ (62). Fig-\ref{fig:C0}(b) shows the orthorhombic crystal structure of CRO sample in 3D and 2D consisting of corner-sharing RuO$_6$ octahedra.

\begin{table*}
 \centering
 \caption{Structural details of CRCO samples, where Avg Ru-O2 and Avg Ru-O1 is average bond length along in-plane and out-of-plane respectively:-}
\begin{tblr}{colspec={|ccccccccc|},hlines}
    {x\\(Sample)} & {Lattice\\constant($\AA$)}&{Volume\\($\AA^3$)} & {a/c} & {Avg Ru-O2\\($\AA$)}& {Avg Ru-O1\\($\AA$)}& {$\phi$ \\Ru-O2-Ru($^\circ$)} & {$\theta$\\Ru-O1-Ru ($^\circ$)}& {(90-$\Omega$)/2\\($^\circ$)}\\
   {0\\(CRO)} & {a = 5.528\\ b = 7.664\\c = 5.361}&227.093& 1.031 & 2.003 & 1.990& 146.722 & 150.307& 9.2 \\
   {0.15\\(CRCO15)} & {a = 5.497\\ b = 7.644\\c = 5.358}&225.128& 1.025 &1.988 & 1.956 & 148.625& 156.345 &4.8\\
\end{tblr}
\end{table*} 
 The lattice constant a, b, and c decrease by 0.54\%, 0.26\%, and 0.07\% respectively with increasing Cr content up to 15\% as shown in Fig-\ref{fig:C0}(c). The decrease in the lattice constant is attributed to the smaller cationic radius of Cr$^{6+}$(0.044 nm)/Cr$^{3+}$(0.0615 nm) (valency of Cr is confirmed by XPS, see next section) than the Ru$^{4+}$ (0.062 nm). Thus increasing the Cr-content squeezes the volume of the unit cell as shown in Fig-\ref{fig:C0}(d). It is noteworthy to mention that the orthorhombicity factor defined by the a/c ratio decreases which indicates that the structure prefers to move toward the tetragonal with Cr doping as shown in Fig-\ref{fig:C0}(d). The refinement indicates that the RuO$_6$ octahedra is distorted and possess tilt along both the equatorial (Ru-O1-Ru) and axial (Ru-O2-Ru) directions. The RuO$_6$ octahedra also possesses in-plane rotation ($\Omega$). The octahedral bond lengths are denoted as a$_{oct}$, c$_{oct}$ and b$_{oct}$, along a, c, and b-axes respectively. The octahedral tilt ((180 - $\theta)/2)$ is related to the octahedral bond lengths by the relation: $c/a =[c_{oct}/a_{oct}][cos((180 - \theta)/2)$ \cite{lu2013role}. Hence in an orthorhombic system, a finite change in tilt angle changes the a/c factor. Octahedral tilt decreases significantly upon Cr doping. The structural distortion (defined as $\Delta = \frac{1}{6}\Sigma^6_{i=1}
\Big[\frac{|d_i - d_{mean}|}{d_{mean}}\Big]^2$, here $d_{mean}$ and $d_i$ denote the average and $i^{th}$ Ru/Cr-O bond length respectively) increases from 1.2$\times$10$^{-5}$ to 2.1$\times$10$^{-4}$ ($\sim$ 18 times) with 15\% Cr-doping. The structural parameters are tabulated in Table II. \\
\subsection{X-ray photoelectrons spectroscopy-}
\begin{figure}
\begin{flushright}
 \includegraphics[width=\linewidth]{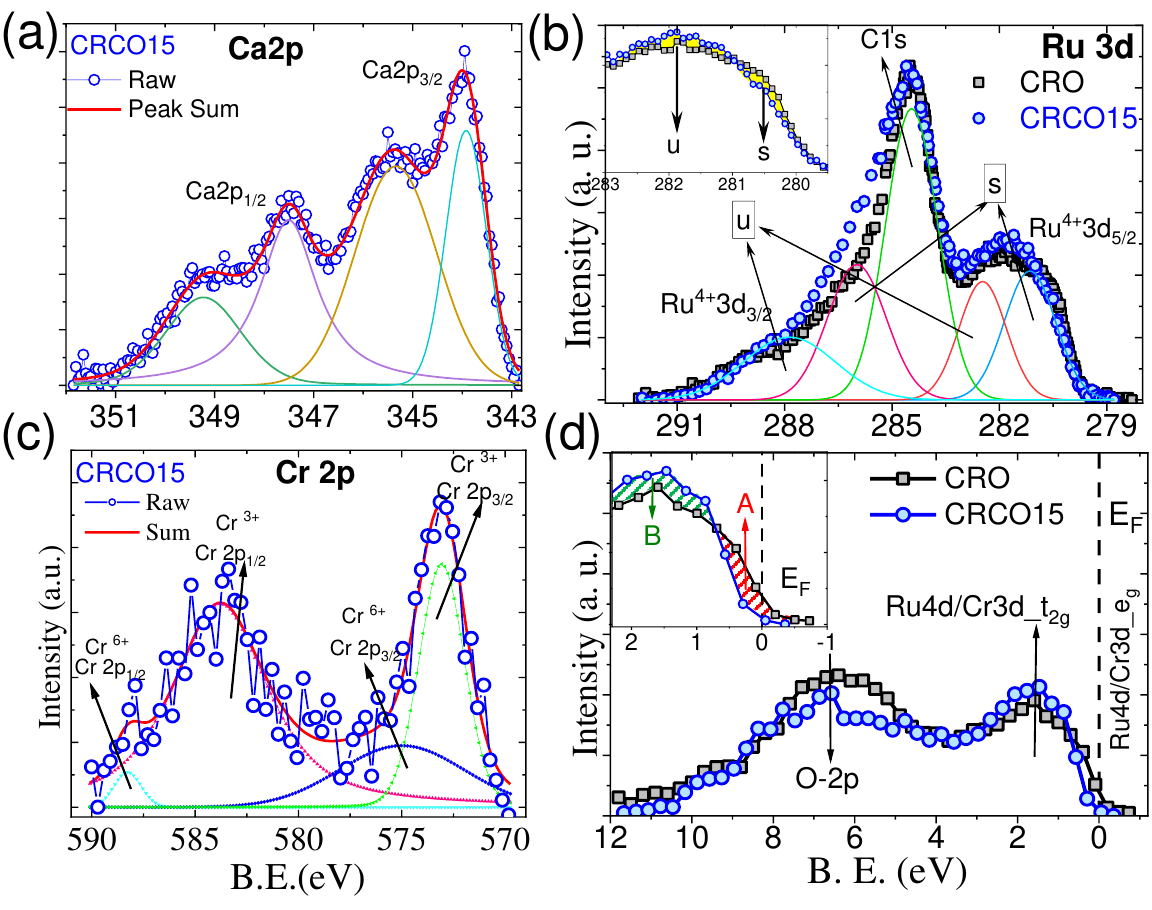} 
 \caption{\justifying{\small{Core spectra of (a) Ca-2p (b) Ru-3d and (c) Cr-2p. The open circles are raw data, red solid lines are the peak sum and different color lines are corresponding fitting peaks. The data   are Shirley background subtracted. (d) XPS valence band spectrum of CRCO15 in comparison with CRO. Inset shows the zoomed-in-view of the valance band near E$_F$, where the red and green shaded region is corresponding to feature-A and feature-B.}}}
\label{fig:C3a}
\end{flushright}
\end{figure}

X-ray photoelectron spectroscopy (XPS) is employed to determine the valence states of the individual constituents as well as the density of state (DOS) at the Fermi level (N(E$_F$)). The XPS survey scan (not shown) confirms the presence of Ca, Ru, Cr, and O atoms. The peaks are analysed by using \href{https://xpspeak.software.informer.com/4.1/} {XPSPEAK-4.1} \cite{yang2006carbon}. The Ca-2p spectra (Fig-\ref{fig:C3a}(a)) are fitted with two spin-orbit doublet (SOD) peaks, Ca2p$_{3/2}$ and Ca2p$_{3/2}$ confirming the presence of Ca$^{2+}$ valence state \cite{ito2008microstructure}. Fig-\ref{fig:C3a}(b) shows the normalized core level Ru-3d XPS spectra for CRO and CRCO15 samples. The Ru 3d SOD ($\Delta E_{SO}^{Ru}$ $\sim$ 5.3 eV) \cite{morgan2015resolving} as: Ru 3$d_{5/2}$ and Ru 3$d_{3/2}$ suggests Ru$^{4+}$ valence state. Each SOD peak breaks into two peaks corresponding to a screened ($s$) and unscreened ($u$) peak in metallic ruthenates due to the screening of core electrons. The screening peak arises (at $\sim$ 2 eV lower binding energy) due to the presence of free electrons (it vanishes in the insulating system like Mott-insulator \cite{kim2004core}). The presence of $s$ and $u$ peaks in all samples suggests the metallic nature of CRCO at room temperature. In CRCO15 the spectral weight of $s$ peak is transferred towards the $u$ peak (at higher binding energy) compared to CRO as depicted in the inset Fig-\ref{fig:C3a}(b). This spectral transfer supports the non-metallic nature of CRCO15 \cite{brar2023lattice}. Fig-\ref{fig:C3a}(c) shows the Cr-2p core level spectra of CRCO. The SOD Cr2p$_{1/2}$ and Cr2p$_{3/2}$ are further split into two, corresponding to two different oxidation states namely Cr$^{+3}$3d$^3_{t_{2g}}$ (magnetic) and Cr$^{+6}$3d$^0_{t_{2g}}$ (non-magnetic) \cite{ranjan2009magneto}. The integrated intensity ratio of Cr$^{3+}$ and Cr $^{6+}$ peak is found to be $\sim$ 2:1 in CRCO15. The overall valence state of Cr follows the relation Cr$^{3+}_{x-y}$Cr$^{6+}_y$ = Cr$^{4+}_x$. It maintains an overall +4 oxidation state at the Ru site (effectively Ru$^{4+}$ substituted by Cr$^{4+}$). The detailed valence structure of CRCO can be written as Ca$^{2+}$Ru$^{4+}_{1-x}$Cr$^{3+}_{x-y}$Cr$^{6+}_y$O$_3^{2-}$, where Cr$^{3+}_{x-y}$Cr$^{6+}_y$ = Cr$^{4+}_x$. The valence state details, binding energy (B.E), and spin-orbit splitting energy ($\Delta E_{SO}$) of the Ca, Ru, and Cr are tabulated in Table III. Fig-\ref{fig:C3a}(d) displays the comparison of valence band (VB) spectra of CRO and CRCO15. The peak near the Fermi edge ($\sim$ 1 eV) is caused by Ru-4d/Cr-3d electrons and the peak around 6.5 eV is caused by O-2p electrons, which indicates a strong hybridization between O-2p and Ru-4d/Cr-3d \cite{maiti2006role}. The intensity of the VB spectra is proportional to the DOS. The VB-spectra fairly matches well with the theoretically calculated DOS spectra \cite{farahani2014valence,vidya2004magnetic}. Finite N(E$_F$) also supports the metallic nature at room temperature. The inset Fig-\ref{fig:C3a}(d) shows the zoomed-in view of VB-spectra near E$_F$, where feature-A and B are attributed to the coherent (delocalized) and incoherent (localized due to electron-electron correlation) electrons respectively \cite{singh2007manifestation}. In CRCO15 (near $\sim$ 0.6 eV) the decrease in the intensity at feature-A is compensated by an increase at feature-B in comparison to CRO. This spectral weight transfer from the lower to the higher binding energy indicates the decrement of delocalized/itinerant electrons with Cr doping. Thus Cr doping supports the nonmetallic ground state of CRCO. In addition, Cr doping also creates disorder and reduces the bandwidth ($W \sim$ 1/N(E$_F)$ \cite{durairaj2006highly}). The reduction of W is confirmed from reducing $\Delta E_{SO}^{Ru}$ values with increasing Cr-doping (see Table III). In short, the XPS study reveals an increase in the localization and disorder with Cr-doping which is also reflected in the magnetic and transport measurements.\\
\begin{table*}
 \centering
 \caption{XPS details: consists Binding and spin-orbit splitting  ($\Delta E_{SO}$) energy of different ions, for C1s peak fixed at 284.5 eV.}
  \begin{tblr}{colspec={|cccccccccc|},hlines}
{Sample} & {Ru3d$_{5/2}$\\(eV)}& {Ru3d$_{3/2}$\\(eV)}& {$\Delta E_{SO}^{Ru}$\\(eV)}& {Ca2p$_{3/2}$\\(eV)}& {$\Delta E_{SO}^{Ca}$\\(eV)}& {Cr2p$_{3/2}$\\(eV)}& {Cr2p$_{1/2}$\\(eV)}& {$\Delta E_{SO}^{Cr}$\\(eV)}& {Species\\($\%$)}\\ 
 {CRO} & {280.9\\ 282.35}&{286.2\\ 288.2} &{5.3\\ 5.85} & {343.9\\345.4} &{3.5\\3.7}& {-\\-} &{-\\-}& {-\\-} & {-\\-}\\
 {CRCO15}&{ 281.5\\282.5}& {286.5\\ 288.1} &{5.0\\5.6} & {343.95\\345.4} & {3.6\\4.1} & {574.90\\577.80} &{584.70\\587.30}& {9.8\\ 9.5} & { 66\% (Cr$^{3+}$)\\ 34\% (Cr$^{6+}$)}\\
\end{tblr}
\end{table*}
\subsection{Magnetic measurement-} 
\begin{figure}
\begin{flushleft}
 \includegraphics[width=1.03\columnwidth]{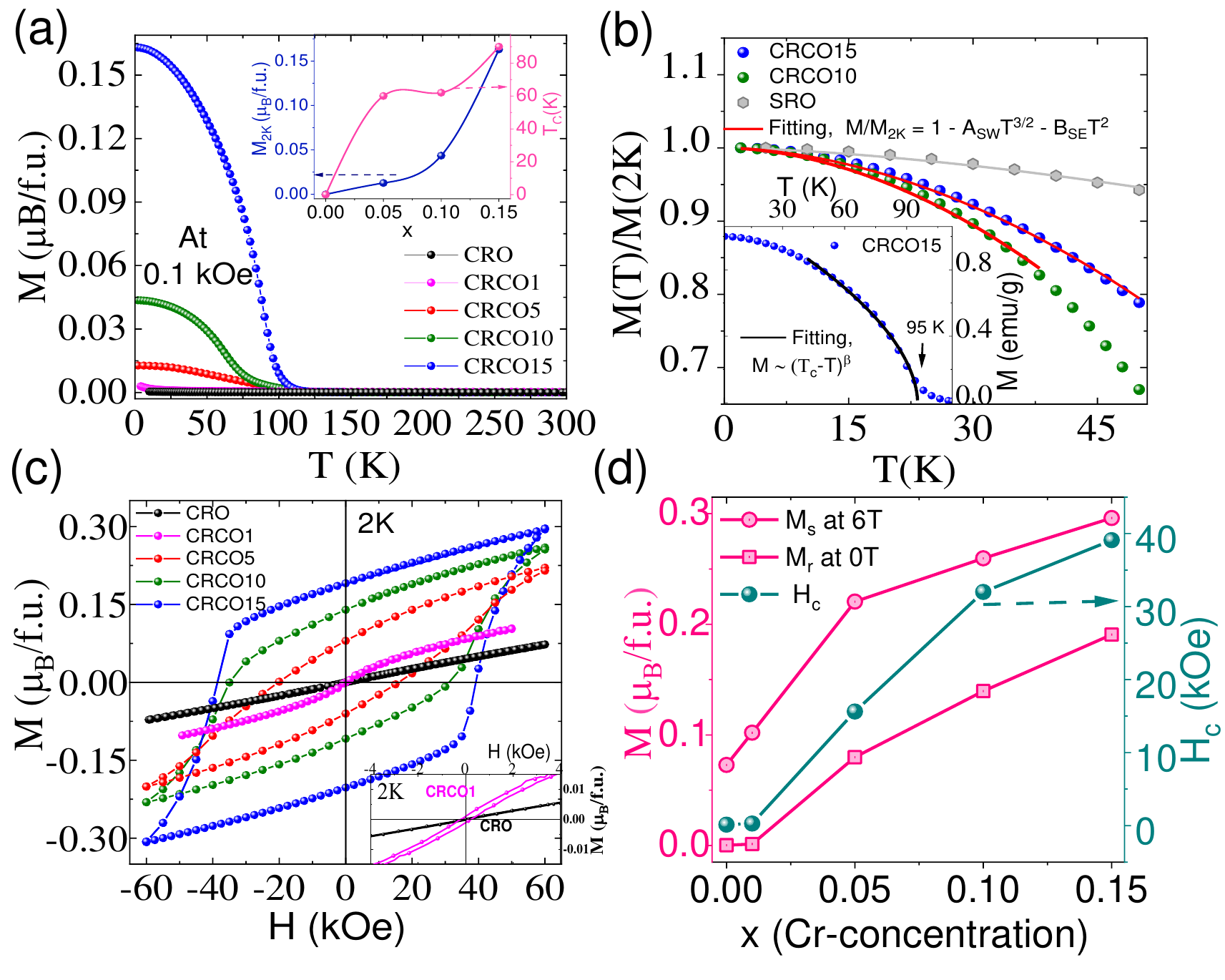} 
 \caption{\justifying{\small{(a) Field-cool M(T) curve measured at 100 Oe of CRCO. Inset shows the variation of T$_c$ and M$_{2 K}$ with different Cr-concentration. (b) M(T)/M$_{2 K}$ curves of CRCO10, and CRCO15 below 50 K. The Inset is M(T)/M$_{2 K}$ vs T of CRCO15, in the critical region (2 - 100 K), solid lines are corresponding fittings of magnetization, Here the SRO data (gray) is plotted for comparison. (c) M-H hysteresis of CRCO at 2 K. Inset shows the zoomed-in M-H data of CRCO1 and CRO. (d) The variation of $M_s$, $M_r$, and $H_c$, with increasing Cr-concentration. }}}
\label{fig:C3d2} 
\end{flushleft}
\end{figure}

The XPS study confirms that Cr exists in two different valence states, Cr$^{+3}$3d$^3_{t_{2g}}$ (magnetic) and Cr$^{+6}$3d$^0_{t_{2g}}$ (non-magnetic). Naturally, one gets magnetic disorder while substituting the Ru with Cr. Thus it is crucial to look into the evolution of magnetization in CRCO. Fig-\ref{fig:C3d2}(a) shows the temperature (T) dependent field cooled (FC) magnetization curve measured with the applied magnetic field of 100 Oe. CRO shows PM-like behavior down to 2 K with AFM fluctuations (as Weiss temperature $\sim$ 176 K), while the others show the gradual growth (from short range to long range) of ferrimagnetism (FE) with Cr-doping. Inset Fig-\ref{fig:C3d2}(a) shows that both T$_c$ and the value of magnetic moment at 2 K (M$_{2 K}$) increase with increasing Cr-concentration, indicating the evolution of FE. The ordering temperature is determined from scaling law fit (M(T)$\sim$(T$_c$ - T)$^\beta$) of magnetization data measured at low field, near the critical region as shown in the inset Fig-\ref{fig:C3d2}(b). It has to be noted that T$_c$ evaluated from M-T scaling law often does not give true value if the sample possess short range magnetic order. The critical exponent $\beta$ $\sim$ 0.55 indicates the mean-field type FM behavior in CRCO15. 
\begin{table}[b]
\begin{center}
\caption{\justifying{Details of fitting parameters of M(T)/M$_{2 K}$ curve at low temperature and near critical regions.}}
\begin{tblr}{colspec={|cccccc|},hlines}
Sample & {A$_{SW}$\\(K$^{-3/2}$)} &{B$_{SE}$\\(K$^{-2}$)}&{J/k$_B$} &{T$_c^{cal}$\\(K)}&$\beta$   \\ 
CRCO5& 1.94E-4&8.77E-5&22.5&116&0.59\\
CRCO10 &1.81E-4&8.44E-5&23.6&121&0.58\\ 
CRCO15&1.72E-4&5.76E-5&24.4 &126&0.55 \\ 
SRO&1.3E-4&2.3E-6&28.7 & 160&0.5 \\ 
\end{tblr}
\end{center}
\end{table}
\begin{table*}
 \begin{flushleft}
 \caption{List of the magnetic parameters obtained from M-H hysteresis and susceptibility fitting. Here S$_{eff}$ is effective free spin.}
\begin{tblr}{colspec={|ccccccccccccc|},hlines}
Sample&{T$_P$\\(K)}& {T$_c$\\(K)}& {H$_c$\\(kOe)} &{M$_s$ ($\mu_B$\\/f.u.)} & {M$_r$($\mu_B$\\/f.u.)}& {M$_{tot}$($\mu_B$\\/f.u.)} &{C (emu K \\/$mol/Oe$)} & {$\mu_{eff}^{Exp}$\\($\mu_B$/f.u.)} & {$\mu_{eff}^{Cal}$\\($\mu_B$/f.u.)}& S$_{eff}$ & {$\theta_{CW}$\\(K)}& $\lambda_G$ \\
  CRO& -&- & -  & 0.07 &-& - & 0.97 &2.793&2.828&1 &-176&-\\
  CRCO5&24 & 70 & 15&0.22 &0.05 & 0.27&0.71 & 2.419&2.845 &0.8& -42&0.96 \\
 CRCO10&58 & 75 & 32 & 0.26& 0.14 & 0.37& 0.81& 2.42&2.862&0.81 &17& 0.93 \\
 CRCO15&90& 95 & 38  & 0.30 &0.19 & 0.41& 0.70 & 2.485&2.878&0.83& 50 &0.78\\
\end{tblr}
 \end{flushleft}
\end{table*}
The collective spin wave excitations (magnon) and single particle Stoner excitation reduce the magnetization at low temperatures through Bloch's T$^{3/2}$ and T$^2$ law respectively. Fig-\ref{fig:C3d2}(b) shows the normalised FC magnetization (M(T)/M(2 K)) below 50 K of CRCO10 and CRCO15. For comparison, an isostructural itinerant ferromagnetic system (T$_c$ = 165 K) SrRuO$_3$ is included. The magnetization data below the critical region fitted with the equation: M(T)= M(2 K)[1 - A$_{SW}$ T $^{3/2}$ - B$_{SE}$ T$^2$], where A$_{SW}$ is spin-wave stiffness constant and  B$_{SE}$ is stoner excitation parameter. A$_{SW}$ is related to the exchange interaction ($\emph{J}$) as: A$_{SW}$ = (0.0587/S)(k$_B$/2JS)$^{3/2}$, and also with T$_c$ as: (A$_{SW}$ )$^{-2/3}$= 2.42 T$_c$ (for Ru$^{4+}$ \cite{snyder2019critical}). We have calculated the J, and T$_c$ by using these relations. The obtained value of A$_{SW}$ , B$_{SE}$, J, T$_c$ and $\beta$ are tabulated in Table-IV. Both A$_{SW}$ and B$_{SE}$ reduce with increasing Cr-content going towards the values for SRO (listed in Table IV), which indicates the possibilities of forming a long-range order at x $\geq$ 0.15. A careful inspection reveals that the low-temperature M-T curve of CRCO reduces faster than that of the SRO, which indicates the presence of larger magneto-crystalline anisotropy (MCA) than SRO. Thus M-H hysteresis study in CRCO is important to elucidate this fact. Fig-\ref{fig:C3d2}(c) shows the M-H hysteresis measured at 2 K for CRCO samples. CRO exhibits a non-hysteretic character resembling a PM-like ground state while the other three samples show large hysteresis with non-saturating behavior. It is noteworthy to mention that tiny (x = 0.01, named CRCO1) Cr doping changes the magnetic ground state from PM-like to FE with small hysteresis ( coercive field (H$_c$) = 200 Oe) as shown in the inset of Fig-\ref{fig:C3d2}(c)). However, the M(T) data of CRCO1 does not show any long-range order down to 4 K. With increasing Cr-doping the coercivity increases and attains a maximum value of 38 kOe for CRCO15, which is 10 times larger than H$_c$ of SRO (3.5 kOe) \cite{sow2012structural}. The high coercivity implies that the sample possesses high anisotropy. Fig-\ref{fig:C3d2}(d) shows the variation of T$_c$, T$_P$, H$_c$, remnant magnetization (M$_r$), and saturation magnetization (M$_s$) at 60 kOe with Cr-doping. These values (tabulated in Table V) increase with Cr doping suggesting the gradual evolution of FE. 
\begin{figure}
\begin{center}
 \includegraphics[width=\columnwidth]{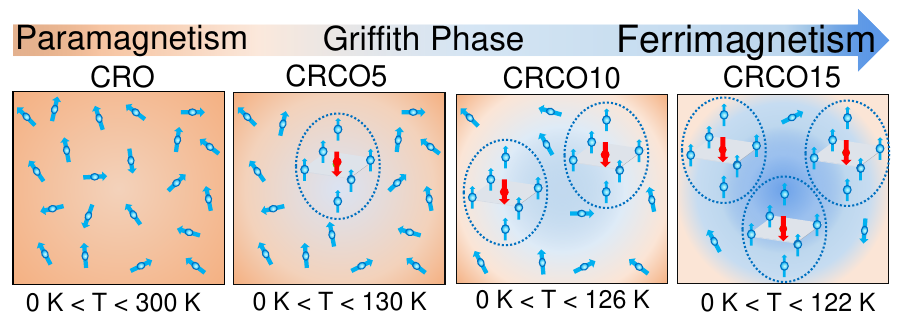} 
 \caption{\justifying{\small{Schematic picture: CRO is assumed to be PM (orange region) where Ru-spin (blue) is not exchange coupled. The short-range FE (blue) cluster region starts to develop with Cr-doping in the PM background. With very small Cr-doping (x $\leq$ 0.05) the cluster can be treated almost \enquote{isolated}. However, for x $>$ 0.05 the clusters start interacting with each other and eventually give a long-range FE-ordered state in CRCO15 at low temperatures. These schematics describe about the evolution of a ferrimagnetic order in paramagnetic CaRuO$_3$ through an intermediate Griffith Phase below the mentioned temperature range.}}}
\label{fig:C3d4} 
\end{center}
\end{figure}
M$_s$ at 2 K for CRCO15 is 0.3 $\mu_B$/f.u. (5 times smaller than that of SRO) which is 15\% of the total spin only moment of Ru$^{4+}$ (2 $\mu_B$/f.u., S = 1) suggesting a weak Ru-Ru exchange. It has to be noted that the Cr-Cr (AFM, T$_N$ = 290 K) exchange is not favorable at low Cr-concentration as reported by \emph{A. Maignan et al.} \citep{maignan2006ferromagnetism}. Hence most of the magnetic signal originates from Ru$^{4+}$-Cr$^{3+}$  exchange interaction. The magnetization can be explained by the seven-atom (6 Ru + Cr) cluster depicted in Fig-\ref{fig:C3d4} (encircled region). In such a cluster each Cr$^{3+}$ ion lies at the center and is surrounded by six nearest neighbour (nn) Ru$^{4+}$ ions. The Ru$^{4+}$-O-Cr$^{3+}$ exchange is AFM \cite{kasinathan2006electronic} whereas Ru-O-Ru is FM (within the cluster). Therefore cluster is ferrimagnetic. Using the chemical formula (Ca$^{2+}$Ru$^{4+}_{1 - x}$Cr$^{3+}_{x - y}$Cr$^{6+}_y$O$_3^{-2}$) the net magnetic moment of each cluster (M$_{cl}$) can be calculated as M$_{cl}$ = 6 M$_{Ru}$ - M$_{Cr}$, where M$_{Ru}$ = g $\mu_B$(1 - x)(S)/(1 - x) = 2 $\mu_B$, for  S = 1, considering a Landé factor (g) equal to 2 and M$_{Cr}$ = g $\mu_B$[(x - y)S]/x = 3$\mu_B$(x - y)/x, for S = 3/2. The total magnetization is M$_{tot}$ = P$\times$ M$_{cl}$, where P = [(1 - x)$^6$(x - y)]/(1 - y) is the total number of clusters (is the probability of Cr$^{3+}$ ion to be at the center of FE cluster). For CRCO15 (x = 0.15, y = 0.33x) the calculated magnetization value is 0.39 $\mu_B$/f.u. Also, there are 25\% [(0.85 - 6 (x - y))100] Ru ions not involved in the cluster (as $\sim$ 30\% Cr$^{6+}$ ions are nonmagnetic), which will also contribute paramagnetically (0.02 $\mu_B$/f.u (PM moment of CRO)). Hence the overall calculated magnetization is 0.41 $\mu_B$/f.u. The experimental value of magnetization for CRCO15 is 0.3 $\mu_B$/f.u. at 60 kOe. This deficit (0.41 - 0.3 = 0.11 $\mu_B$/f.u.) in the saturation magnetization has two possible reasons: (i) the presence of a large anisotropy field, or (ii) all Cr$^{3+}$ ions are not participating in seven-atom clusters. The value of M$_{tot}$ calculated for other CRCO samples are also tabulated in Table V. \\
 \begin{figure}
\begin{flushleft}
 \includegraphics[width=\linewidth]{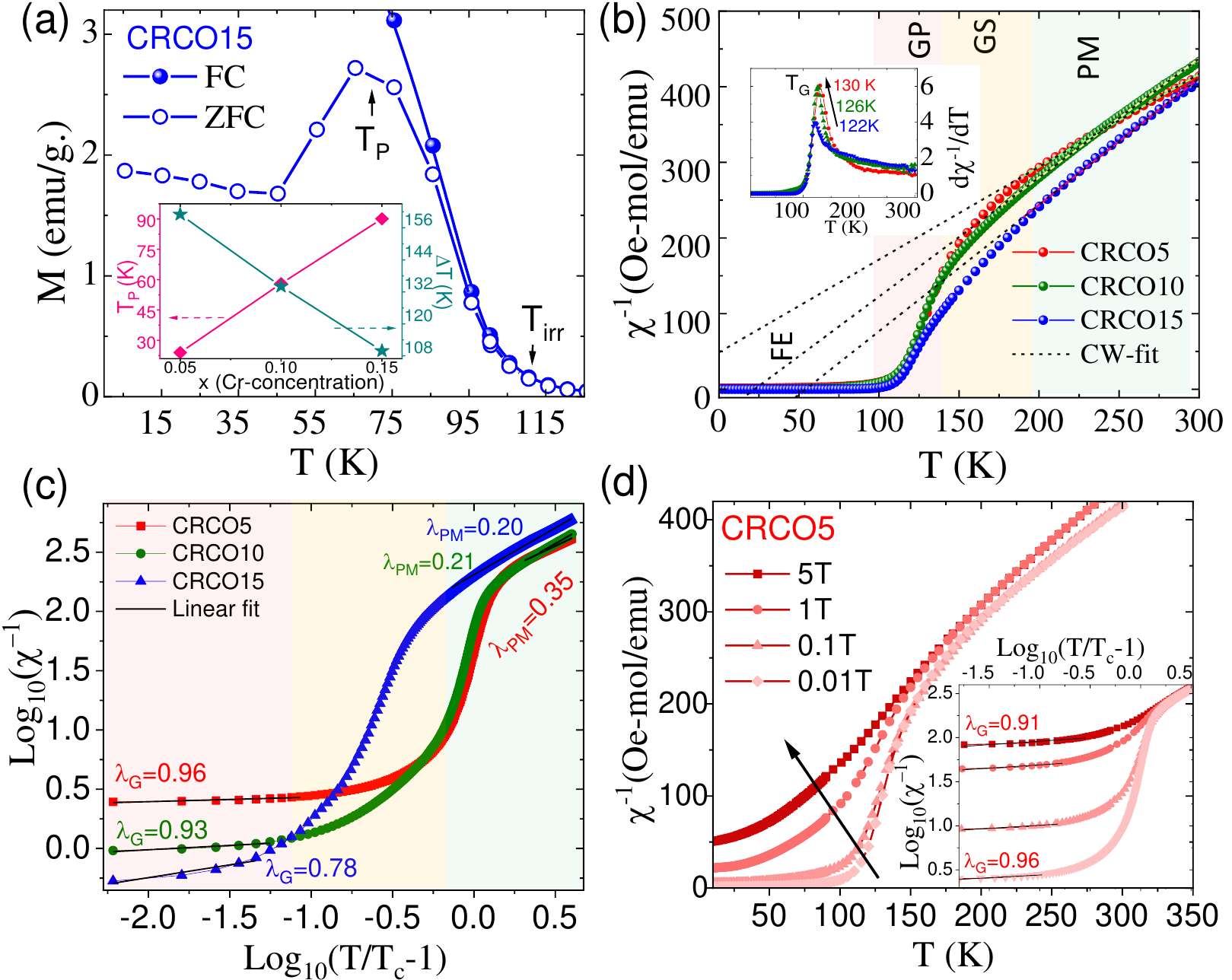} 
 \caption{\justifying{\small{(a) Zoomed in view of ZFC and FC magnetization data of CRCO15. FC-ZFC curves bifurcate at T$_{irr}$. Inset shows the variation of T$_P$ and $\Delta$T with Cr-content. (b) $\chi^{-1}$ vs T data of CRCO samples dashed line is the Curie-Weiss fitting in PM region. Inset shows the d$\chi^{-1}$/dT vs T plots, where the peak at T$_G$ is Griffiths temperature. (c) Log ($\chi^{-1}$) vs Log (T/T$_c$-1) plot, where dashed lines show the modulated Curie-Weiss or power low ($\chi^{-1} \propto$ (T-T$_c$)$^{1-\lambda}$) fitting in PM and GP region. (d) $\chi^{-1}$ vs T data of CRCO5 at different applied magnetic field. Inset is power low fitting at different fields.}}}
\label{fig:C4b} 
\end{flushleft}
\end{figure} 
A careful inspection of the FC and zero fields cool (ZFC) magnetization data of CRCO15 (Fig-\ref{fig:C4b}(a)) indicates that the bifurcation temperature (T$_{irr}$) is much above the ZFC peak temperature (T$_P$). This hints at the presence of magnetic clusters \cite{bray1987nature,magen2006observation,pramanik2010griffiths}. Inset of Fig-\ref{fig:C4b}(a) shows the variation of T$_P$ and separation $\Delta$T (= T$_{irr}$ - T$_P$) with Cr-doping. The value of $\Delta$T reduces monotonically with increasing Cr-doping suggesting a gradual evolution of a magnetic order. To verify whether CRCO possesses spin glass \cite{mydosh1993spin} ground state or any other slow dynamics, the frequency-dependent AC-susceptibility measurements (not shown) are done. We did not find any frequency dependence of the magnetic ordering temperature, confirming the absence of a spin glass ground state. Further study of the field cool dc-susceptibility data reveals non-linearity in the inverse susceptibility ($\chi^{-1}$) above T$_c$ as shown in Fig-\ref{fig:C4b}(b). Below T$_c$ the system has FE ordered phase and above 200 K it exhibits a regular PM phase. In the intermediate temperature range (140 $\leq$ T $\leq$ 200 K) it undergoes a disordered phase, where the values of magnetic exchange are assigned randomly at different lattice sites because of local disorder within the crystallographic structure known as Griffiths singularity (GS) \cite{bray1982eigenvalue}. The susceptibility diverges in the 90 to 140 K temperature range. In this particular case, GP developed as a result of competing magnetic interactions of PM (free Ru), short-range FM (Ru-Ru), and AFM (Ru-Cr). The non-linear downturn in $\chi^{-1}$ observed for CRCO samples is a typical characteristic of the GP. It is important to note that the non-linearity in $\chi^{-1}$ decreases with Cr content. The temperature at which the downturn starts is marked as Griffiths temperature $T_G$ calculated from the $d\chi^{-1}/dT$ curve, as shown in the inset of Fig-\ref{fig:C4b}(b). In PM region the $\chi^{-1}$ follows the Curie-Weiss law: ($\chi$ = $\chi_0$ +C/(T-$\theta_{CW}$), where $\chi_0$ is Pauli susceptibility, $C$ and $\theta_{CW}$ is Curie constant and Curie-Weiss temperature respectively. The dashed lines in Fig-\ref{fig:C4b}(b) show the Curie-Weiss Law fitting. By using the obtained value of the Curie constant, the effective PM moment ($\mu_{eff}$) can be calculated as- $\mu_{eff}$= 2.828 (C)$^{1/2}$. The experimental and theoretical values of $\mu_{eff}$ are tabulated in Table V. In case of CRO the difference between $\mu_{eff}^{cal}$ and $\mu_{eff}^{exp}$ is negligible whereas in CRCO the difference is large ($\sim$ 0.4\%). This indicates that some of the magnetic moments of CRCO in the PM region are correlated (not free). They form clusters in the PM region. The strength of the GP explicitly characterised by relation: $\chi^{-1} \propto$ (T-T$_c$)$^{1-\lambda}$, where 0 $\le$ $\lambda$ $\leq$ 1 is susceptibility exponent. Fig-\ref{fig:C4b}(c) shows the log ($\chi^{-1}$) vs log (T/T$_c$-1) plot. The fitted values of $\lambda_G$ (in GP regime) and $\lambda_{PM}$ (in PM regime) are listed in Table-V. A large value of $\lambda_G$ (0.96) for CRCO5 demonstrates that strong GP exists in a low Cr doped sample (however x = 0 and 0.01 do not show any signature of GP). Further confirmation of GP is done by measuring susceptibility at various applied fields ($H$) as shown in Fig-\ref{fig:C4b}(d). The nonlinearity in $\chi^{-1}$ reduces (softening of downturn) with increasing field from 0.01 T to 5 T (indicated by arrow) is a true signature of the GP. The exponent $\lambda_G$ reduces (shown in the inset of Fig-\ref{fig:C4b}(d)) with increasing field. It has to be noted that $\Delta$ (defined in structural section) can also influence the Griffiths instabilities \cite{lu2013role,banik2018evolution}. In this case, it is observed that $\lambda_G$ decreases as $\Delta$ increases. In Fig-\ref{fig:C4b}(b)) and (c) data do not merge at high temperatures (although it is expected to merge  \cite{magen2006observation}). Such parallel shift may arise due to various reason: competition between Pauli paramagnetic susceptibility ($\chi_{P}$) and Curie-Weiss susceptibility, instrumental background which contributes significantly at very high field. However, for low field measurements (0.01 T and 0.1 T) in Fig-\ref{fig:C4b}(d), the susceptibility data merged at high temperature. The magnetization relaxation measurement (using the protocol reported by \cite{harikrishnan2010memory}) at 5 K for CRCO5 exhibits (data not shown) slow dynamics with characteristic relaxation time $\sim$ 20 sec, which is several orders of magnitude larger than the microscopic spin flip time for canonical spin-glass system.
As we increase the Cr concentration (x $>$ 0.05) the value of $\lambda_G$ decreases, implying a reduction of GP. While a gradual evolution of ferrimagnetic ordering takes place against such short-range Griffiths phase. In this correlated and disordered system, the non-metallic (localized) nature is further elucidated in the transport data.
\subsection{Transport measurements-} 
\begin{figure}
  \begin{flushleft}
 \includegraphics[width=1.02\linewidth]{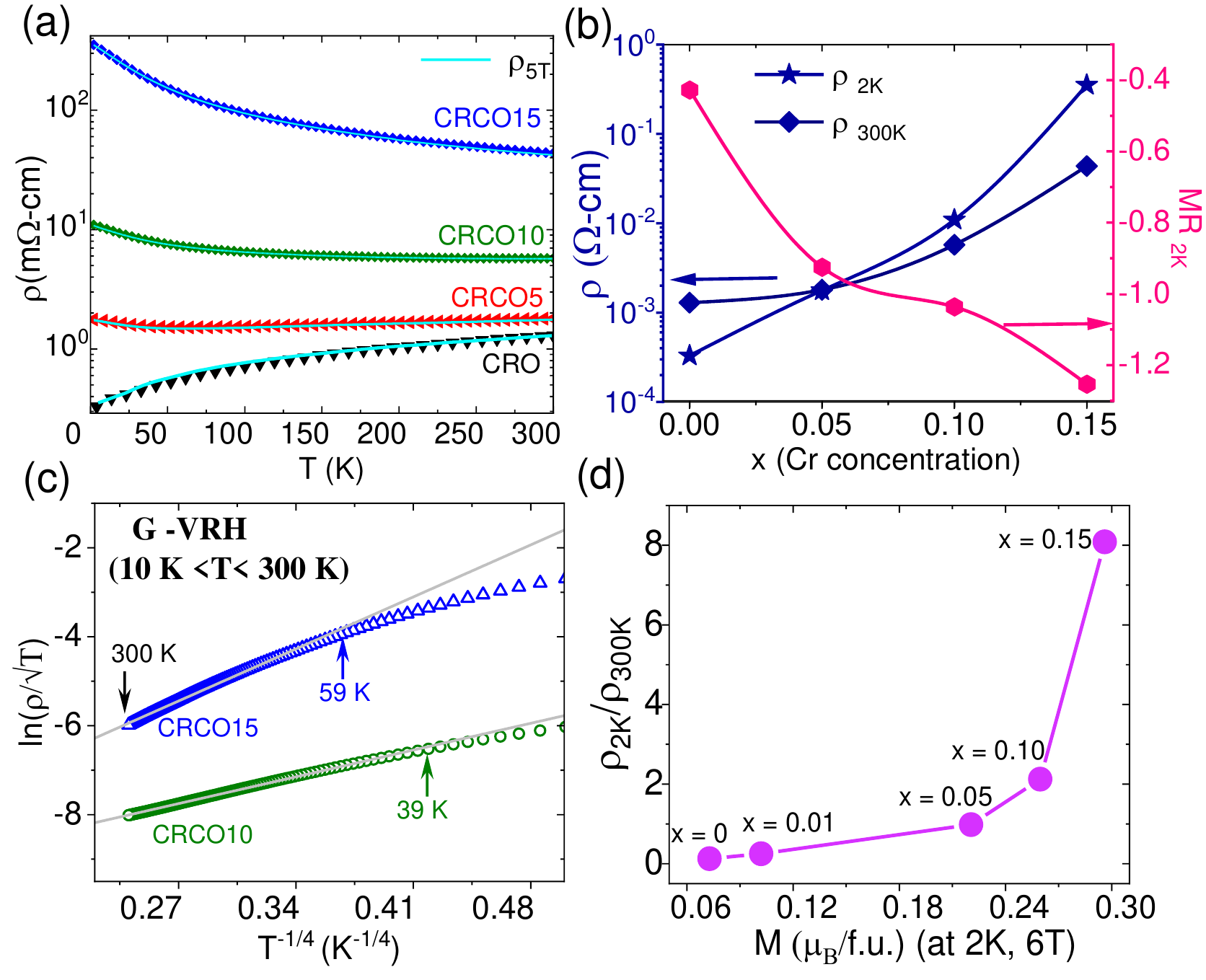} 
 \caption{\justifying{\small{(a) $\rho(T)$ vs T data of CRCO, the continuous line (sky blue) represent $\rho(T)$ at 5 T. (b) Variation of $\rho_{2 K}$, $\rho_{300 K}$ (left y-axis) and MR$_{2K}$ (right y-axis), with increasing Cr-concentration. Resistivity fitting of (c) G-VRH (ln($\rho(T)/\sqrt{T}$ vs T$^{-1/4}$) model for CRCO10 and CRCO15, where solid gray lines are the corresponding linear fitting of resistivity in different temperature ranges. (d) is the variation of resistivity ratio ($\rho(2K)$/$\rho(300K)$) as a function of saturation magnetization (M) at 2K, 6T, for different values of x.}}}
 \label{fig:C6} 
\end{flushleft}
\end{figure}

Fig-\ref{fig:C6}(a) shows the temperature-dependent resistivity $\rho(T)$ vs T graph at 0 T and 5 T for CRCO. CRO exhibits a metallic behavior down to the lowest temperature whereas, CRCO5 holds the metallic nature down to 62 K and exhibits resistivity upturn while lowering the temperature. CRCO10 and CRCO15 are more resistive with a progressively non-metallic nature. At 5 T resistivity curves do not show much change (actually the resistivity reduces by $\sim$ 1\% at 2 K) than that of at 0 T curves. Fig-\ref{fig:C6}(b)(left y axis) shows the variation of $\rho_{300 K}$ and residual resistivity ($\rho_{2 K}$) with increasing the Cr-concentration. In CRCO15, the value of $\rho_{300 K}$ is nearly two orders higher than in CRO. This rise in resistivity with Cr doping is primarily caused by disorder and electron correlation among structural distortion. The role of disorder and correlation on electron transport can be understood by analyzing the temperature-dependent resistivity fit. In the limit x $\geq$ 0.10, the Greaves Variable Range Hopping (G-VRH) or modified Mott-VRH mechanism \cite{greaves1973small,mott1971electronic} follows, which is expressed as:
\begin{equation}
\rho(T) =  \rho_{\infty}\sqrt{T} \exp \Big[\Big(\frac{T_M}{T}\Big)^{1/4}\Big]
\end{equation}
where $\rho_{\infty}$ is a pre-factor term related to phonon density (as the hopping of electron is phonon-assisted) and $T_M$ is the Mott's characteristic temperature. Physically $T_M$ is the measure of the degree of disorder and localization in the system. $T_M$ is related with localization length (1/$\alpha$) and  N(E$_F$) as : $T_M$ = 19.45 $[\frac{\alpha^3}{N(E_F)k_B }]$. The 1/$\alpha$ is the length scale over which the electron wave function is localized. In the case of metals, 1/$\alpha$ is infinitely large but finite in the case of insulators \cite{hine2007localization}. Fig-\ref{fig:C6}(c) shows the resistivity fitting of the G-VRH model. The G-VRH model fits suggest that the role of disorder and correlation both are important. To calculate 1/$\alpha$, we have taken the theoretical DOS of CRO at Fermi energy as N(E$_F$) = 4.1 states/f.u./eV (= 7.22 $\times$10$^{28}$ states m$^{-3}$/eV) \cite{rao2001electronic}. The fitting parameters of G-VRH are tabulated in Table VI. The 1/$\alpha$ value decreases with increasing Cr-doping indicating the presence of more disorder in higher Cr-doped samples. Fig-\ref{fig:C6}(d) displays the variation of resistivity ratio (RR) ($\rho_{2K}$/$\rho_{300K}$) as a function of saturation magnetization (M$_s$) for different values of x. For x $\geq$ 0.1, a large change in RR with a relatively small change in M$_s$ indicate that transport and magnetization process are independent. On the contrary, for x$\leq$ 0.05, a linear change in RR with M$_s$ suggesting that the magnetism and transport are correlated \cite{sarkar2015correlation}. It has to be noted that the orthorhombicity (a/c factor) reduces with Cr-doping even though there is no abrupt structural change (first-order structural change). In short, the ground state of CRCO evolved from an itinerant (metallic) picture towards localized (nonmetallic).\\
\begin{table}[b]
\begin{center}
\caption{\justifying{Details of fitting parameters of resistivity data.}}
\begin{tblr}{colspec={|cccc|},hlines}
Sample & CRCO5 & CRCO10 &CRCO15\\
T$_{M}$(K)& 879 &5464.7&77983\\
1/$\alpha$ (nm)&1.53&0.83&0.34\\ 
\end{tblr}
\end{center}
\end{table}
\begin{figure}
 \begin{center}
 \includegraphics[width=\linewidth]{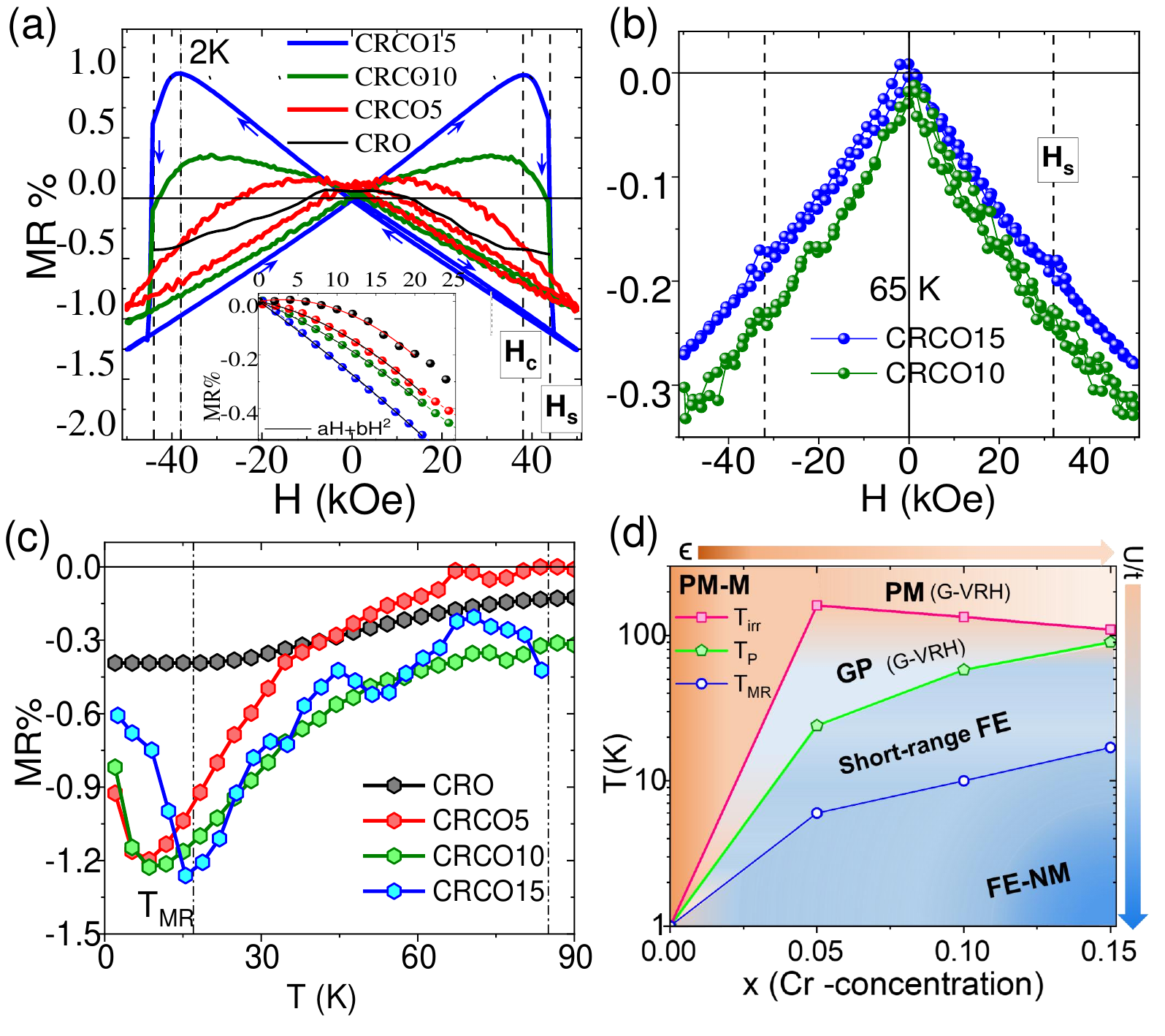} 
 \caption{\justifying{\small{(a) MR plot at 2 K, where dashed lines at 38 kOe and 44 k Oe correspond to H$_c$ and H$_s$ field respectively. The inset displays virgin MR, where the solid black line is aH + bH$^2$ fitting below 20 kOe. (b) MR plot at 65 K, where the dashed line at 32 kOe is at the H$_s$ field. (c) Temperature dependence MR of CRCO samples. (d) T-x phase diagram, shows paramagnetic-metal (PM-M) phase gradually becomes Ferrimagnetic non-metallic (FE-NM) phases with increasing Cr doping and lowering temperature. The horizontal and vertical arrows show the increasing disorder potential ($\epsilon$) and effective Coulomb ($U/t$) energy respectively. The orange colour represents PM-M and the Blue colour represents FE-NM with an intermediate GP (light orange-blue mixed). A short-range ferrimagnetic phase stabilized (light blue region) below T$_P$. T$_{MR}$ (peak in MR vs T plot) represents the temperature at which the maximum spin fluctuations occur where a long-range ferrimagnetic order is likely to be formed.}}}
 \label{fig:C10a} 
\end{center}
\end{figure}
Further, the low-temperature magneto-resistance (MR) measurements are performed down to 2 K with the magnetic field up to 5 T (as shown in Fig-\ref{fig:C10a}(a)). The CRO exhibits weak negative MR ($\sim$ 0.4\%) but doesn't show any hysteretic behavior as a function of the field. Whereas the Cr-doped samples exhibit R-H hysteresis and the hysteretic nature increases with Cr doping. The overall nature of the negative MR resembles fairly well with other established ferromagnetic Ruthenates \cite{maignan2006ferromagnetism,gupta2020electronic,zhang2007competition}. Interestingly, a sharp jump in MR is noticed (for CRCO10 and CRCO15) much above the coercive field (switching field H$_s$ $\sim$ 44 kOe). The possible origin of this sharp jump is attributed to the full alignment of AFM Ru-Cr moments along the magnetic field direction (metamagnetic-like transition). The negative MR increases with increasing Cr-doping and reaches the maximum value (-1.25 \%) for CRCO15 as shown in Fig-\ref{fig:C6}(b) (right y-axis). The inset of Fig-\ref{fig:C10a}(a) shows the virgin curve of MR (0 - 25 kOe) for CRCO measured at 2 K. A non-linear to linear MR trend is evident with increasing Cr-doping. The MR data below 20 kOe is fitted with $a$H + $b$H$^2$ dependence. A sharp increase in the values of $a$ is noticed with Cr\%. For CRO (CRCO15) the values of $a$ and $b$ are 4.2 $\times$ 10$^{-6}$ and 7.2 $\times$ 10$^{-6}$ (2.1 $\times$10$^{-5}$ and 1.2 $\times$10$^{-10}$) respectively. Thus it is evident that the non-linear behavior decreases with increasing Cr-content. Non-linearity in MR can occur due to several factors such as spin scattering (SS) with impurity ion and multiple grain boundary (MGB) effects. Generally, SS is dominant in the itinerant system. As the Cr doping increases, the system evolves towards localised character which makes SS less dominant and gives linear MR in CRCO15. Fig-\ref{fig:C10a}(b) shows the MR data of CRCO10 and CRCO15 at 65 K. The hysteresis in MR almost disappears although the switching of AFM Ru-Cr moments along the magnetic field at 32 kOe is visible. Fig-\ref{fig:C10a}(c) shows the MR vs T data of CRCO samples, which is calculated from resistivity data at 0 and 5 T magnetic field. For CRO the MR remains nearly constant while, for CRCO samples the MR gradually reduces from $\sim$ 85 K (short range FE order) and achieves a negative peak at T$_{MR}$. Such a peak in MR indicates the formation of long-range magnetic order (when the spin fluctuation is expected to be maximum) possibly occurring at a much lower temperature than T$_P$. The value of T$_{MR}$ increases with increasing Cr doping from 8 to 17 K. Finally the essence of this study is corroborated in a phase diagram as shown in Fig-\ref{fig:C10a}(d). This T-x phase diagram depicts a gradual evolution of ferrimagnetic non-metallic (FE-NM) phase from a paramagnetic-metal (PM-M). In the intermediate doping the GP is prominent. At high temperatures, $U$ does not change much with tiny amount of Cr-doping. The competition is essentially between $t$ and $\epsilon$ in Eq-\ref{equ:1}. With the increase in $\epsilon$, the hopping probability gets reduced and system moves towards a non-metallic state. However, in correlated systems especially in Ruthenates, the effective coulomb energy ($U/t$) also increases while lowering the temperature \cite{durairaj2006highly}. Thus the non-metallic nature can be attributed disorder and correlation. The orange colour represents PM-M and the Blue colour represents FE-NM with an intermediate GP (light orange-blue mixed).   
\section{CONCLUSION-} CaRuO$_3$ lies on the verge of magnetic order. Cr-doping acts as a perturbation, which gradually enhances the effective electron correlation in CaRu$_{1-x}$Cr$_x$O$_3$. We have studied the structural, spectroscopic, magnetic, and transport properties of correlated Ruthenate system CaRu$_{1-x}$Cr$_x$O$_3$ (0 $\leq$ x $\leq$ 0.15). All samples preserve the orthorhombic crystal structure with space group $Pnma$ (62) although the orthorhombicity factor (a/c) and the unit cell volume reduce with Cr-doping. XPS study confirms the presence of magnetic (Cr$^{3+}$) and non-magnetic (Cr$^{6+}$) impurity in 2:1. The presence of GP is confirmed from the magnetic measurements. Weak FM Ru-Ru exchange is installed with Cr-doping. The presence of an FE cluster is evident from the ZFC-FC bifurcation much above the ordering temperature. The smaller magnetic moment (0.3 $\mu_B$/f.u.) is explained in terms of seven atom cluster, where Cr is aligned antiferromagnetically with six $nn$ Ru ions. The non-saturating M-H hysteresis and small size of magnetic moment at 6 T (compared to 2$\mu_B$) implies that the overall magnetic structure is FE in nature. The sharp peak in the temperature dependence of MR suggests a long-range magnetic order at $\sim$ 8 to 17 K for CRCO5 to CRCO15. Temperature-dependent resistivity in conjunction with the XPS study reveals an evolution of the nonmetallic state in Cr doped CaRuO$_3$. A new phase diagram is constructed summarising this work. 
\section*{ACKNOWLEDGMENTS-} The authors acknowledge SRG SERB Grants (SRG-2019-001104, CRG-2022-005726, EEQ-2022-000883), India, and Initiation Grant (IITK-2019-037), IIT Kanpur, for financial support. Special thanks to S. Yonezawa (Kyoto University, Japan) for the magnetic and magnetotransport measurements. 

\bibliographystyle{apsrev4-2}
\bibliography{Bib}
\end{document}